\documentclass[twocolumn]{article}

\usepackage{pgfplots, pgfplotstable}
\pgfplotsset{compat=newest}

\usepackage{amsmath}
\usepackage{natbib}
\setcitestyle{numbers}
\usepackage{hyperref}
\usepackage[margin=0.5in]{geometry}

\AtBeginDocument{%
  }

\begin{document}

\title{Generating Finite Element Codes combining Adaptive Octrees with Complex Geometries}

\author{Eric Heisler$^1$ \and Cheng-Hau Yang$^2$ \and Aadesh Deshmukh$^1$ \and Baskar Ganapathysubramanian$^2$ \and Hari Sundar$^1$}

\date{
$^1$University of Utah\\
$^2$Iowa State University\\[2ex]
\today
}

\maketitle

\begin{abstract}
We present a high-level domain-specific language (DSL) interface to drive an adaptive incomplete $k$-d tree-based framework for finite element (FEM) solutions to PDEs. This DSL provides three key advances: (a) it abstracts out the complexity of implementing non-trivial FEM formulations, (b) it simplifies deploying these formulations on arbitrarily complicated and adaptively refined meshes, and (c) it exhibits good parallel performance. Taken together, the DSL interface allows end-users to rapidly and efficiently prototype new mathematical approaches, and deploy them on large clusters for solving practical problems. We illustrate this DSL by implementing a workflow for solving PDEs using the recently developed shifted boundary method (SBM). The SBM requires approximating relatively complicated integrals over boundary surfaces. Using a high-level DSL greatly simplifies this process and allows rapid exploration of variations. We demonstrate these tools on a variety of 2-D and 3-D configurations. With fewer than 20 lines of input, we can produce a parallel code that scales well to thousands of processes.  This generated code is made accessible and readable to be easily modified and hand-tuned, making this tool useful even to experts with the target software.
\end{abstract}

\section{Introduction}
Solving partial differential equations (PDEs) using the finite element method (FEM) involves a linear system assembly process that lends itself well to parallel computation. There are some aspects of the computation that can be greatly simplified with the use of a structured mesh. When limiting the mesh to regularly shaped elements that are either uniformly sized or scaled versions of each other, the calculation of integrals can be greatly simplified. The geometric factors and integration points for each element do not need to be computed or stored. Reference element values can be scaled as needed. 
Another challenge is the efficient partitioning of the problem in a way that balances the load across processors. This task is complicated further when localized refinement is needed. Structured meshes can provide simpler analysis and adjustment of partitioning. The type of structured mesh considered in this work is a $k$-d tree, such as a quadtree in 2-D and octree in 3-D. These offer a very efficient structure that is particularly easy to refine and can be analyzed for load balancing purposes \cite{SundarSampathBiros08,Fernando19,Burstedde11}.

The major drawback to using this kind of structured tree mesh is that they do not conform to irregularly shaped domains. They are ideally suited to axis-aligned rectangular domains, and any irregular geometry will require careful handling of the boundaries. There are a number of techniques that have been developed for this purpose, which are discussed further in the next section. In this work, we focus on the shifted boundary method \cite{main2018shifted}. This method allows us to perform integration over the full structured elements intercepting the boundary while accurately satisfying the boundary conditions defined on the irregular domain. This means that the full benefit of the quasi-structured mesh can be realized while enabling support for a much broader set of applications and domains.

Another important consideration is the need for incomplete trees. Since some of the elements in a full tree may lie completely outside of the irregular domain, those parts of the tree do not contribute to the calculation and should be removed. This irregularity in the tree complicates the mesh generation and load balancing procedures. The tree library used in this study can accommodate these needs, but as discussed below, it can have a substantial cost.

A major contribution of this work is the development of a code generation target module for the domain-specific language, Finch \cite{Heisler23,FinchWebsite}. This DSL supports code generation for external software packages, such as $k$-d tree-based finite element libraries like \texttt{Dendro}. Direct use of \texttt{Dendro} for finite element calculations requires a substantial programming effort and intimate knowledge of the available functionality. To compound this difficulty, the shifted boundary method involves a complicated set of integrals over the boundary that depend on the type of boundary conditions present. Finch greatly simplifies the utilization of this library and allows intuitive input of the integrals in a form resembling the mathematical expressions.

There are several other DSLs that are coupled with parallel finite element libraries~\cite{fenics,mfem,metafem}, but these are generally stand-alone products that are based on a specific set of tools and encapsulate the entire process from user input to solution output. Finch, on the other hand is designed to accept code generation target modules for an external tool set without a direct coupling to those tools. The generated code is also intended to be accessible to the user to allow inspection and modification.

As an example of this benefit to the programmer, we demonstrate a Finch script consisting of less than twenty lines that generates over 2500 lines of C++ code for solving a PDE with the \texttt{Dendro} library. Although a majority of this is statically generated code for setting up the library and defining utility functions, writing these parts by hand would require expertise with the library and substantial effort.
The intuitive interface opens this powerful tool to a wider audience of domain scientists, while keeping the generated code accessible and readable for more advanced users to fine-tune. For examples of input and generated code, see Appendix A and B respectively.

\section{Background}
\subsection{Treating complex geometries using the Shifted Boundary Method}

Standard numerical approaches for solving PDEs on complex geometries usually rely on generating body-fitted meshes. This is a major bottleneck, as creating an analysis-suitable body-fitted mesh with appropriate refinement around the complex geometry is usually time-consuming and labor-intensive. Immersed boundary methods (IBM) alleviate the requirement of body-fitted meshes by relaxing the requirement that the mesh conforms to the object \cite{mittal2005immersed,peskin1972flow}. We consider immersed methods in the context of FEM-based discretizations. There are several immersed methods in use: IBM (Immersed Boundary Method) \cite{peskin1972flow}, IMGA (Immersogemetric analysis) \cite{Kamensky:2015ch}, and FCM (Finite Cell Method) \cite{Dominik2012}. For IMGA and FCM, the Dirichlet boundary conditions are enforced by Nitsche's method on the exact positions of the geometries. However, cut-cells in IMGA and FCM (i.e., cut elements that intersect the physical domain on small volumes) can deteriorate the conditioning of the stiffness matrix in the finite element method. The SBM (Shifted Boundary Method) \cite{main2018shifted} was recently proposed in the family of immersed finite element methods. Instead of enforcing the boundary conditions at the exact place on the geometries, the SBM shifts the location where one enforces the boundary conditions from the true boundary to a surrogate boundary consisting of the nearby mesh edges or faces to prevent the cut-cell issues of FCM and IMGA. SBM has already been applied in various fields like fluid simulations \cite{MainS18a}, structure simulations \cite{atallah2021shifted}, and free surface flow \cite{osti_1851595}. In addition to its use with unstructured triangular and tetrahedral meshes \cite{main2018shifted,MainS18a}, SBM has been recently been applied in an octree-based finite element framework \cite{saurabh2021scalable}. The principal idea of SBM is to shift the boundary condition applied on the true boundary $\Gamma$ to a surrogate boundary $\Tilde{\Gamma}$ using a Taylor series expansion. This produces an additional set of three terms in the weak formulation that involve surface integration over the surrogate boundary. Additionally, these surface integration expressions require non-trivial distance calculations from a surface Gauss-point on the surrogate boundary to the true boundary. These terms are discussed in detail for various equations -- Poisson, Linear Elasticity, and Navier Stokes -- in the last section of this paper.

\subsection{$k$-d tree generation for complex geometries}
We consider two and three dimensional trees, quadtrees and octrees respectively, though the concepts apply more generally. The tree generation tool we are using is based on Dendro-kt, which provides balanced, partitioned, parallel tree structures that are useful for large-scale numerical PDE discretizations\cite{Ishii19}. The base library is combined with a set of finite element utilities. This framework has been proven in large-scale parallel simulations~\cite{saurabh2021scalable}.

A feature of particular relevance to this work is the ability to carve out a subdomain from the full tree that is based on a provided irregular geometry description. This geometry can be supplied by a Gmsh mesh file for 2-D or an STL file for 3-D. During the tree construction, elements are flagged as interior, exterior, or intercepted depending on their location relative to the surface of the mesh. The interior and exterior can be swapped depending on whether the mesh corresponds to the domain or a carved out void in the domain. This flag also allows refinement along the boundary of the geometry to better approximate the non-conforming surface. Several examples of this can be seen below in the Demonstrations section.

\section{Domain Specific Language}
Developing a FEM code based on \texttt{Dendro} requires a substantial programming effort and familiarity with this set of tools. 
Finch provides the high-level software interface we used to generate the C++ code for \texttt{Dendro}. 
The Julia-based DSL is intended for numerical solution of PDEs using different discretization methods [reference included in final version], including finite element and finite volume. It uses a variety of code generation targets, including internal Julia options and external frameworks in other languages, that can be chosen to suit the needs of the problem. Finch provides a high-level language that allows simple and intuitive input of the boundary integral terms coming from SBM. See Appendix A for a simplified example input.

A PDE can be specified in Finch by first converting it to a weak form expression in a residual form. 
The following advection--diffusion equation will be used to illustrate the process. Note that in the input the integrals over the volume are implied. To handle other integrals, such as those over the boundary or all element surfaces, those terms need to be labeled as in \texttt{surface( ... )}. This example omits the boundary integral for simplicity.\\
\begin{tabular}{r|l}
PDE & $\frac{du}{dt} - D\nabla^{2}u - \textbf{b} \cdot \nabla u = f$\\
Weak form & $\left( \frac{du}{dt}, v \right) + (D\nabla u , \nabla v) - (\textbf{b} \cdot \nabla u,v) - (f,v) = 0$\\
DSL input & \texttt{Dt(u*v) + D*dot(grad(u),grad(v))}\\
& \texttt{ - dot(b, grad(u))*v - f*v}
\end{tabular}

This input expression will first be expanded into a more verbose symbolic form. Then the time derivative will be discretized according to the specified time integration method. For example, if we specify a simple backward Euler time stepper with\\ \texttt{timeStepper(EULER\_IMPLICIT)}, the term $\frac{du}{dt} * v$ will be transformed into $\frac{1}{dt} (u - u_{old})*v$ where $u_{old}$ represents a known value from the previous time step. Since it is an implicit method, all other terms with the unknown $u$ will remain unchanged. If an explicit method were chosen, the other terms would now involve $u_{old}$.

The final symbolic form is arranged into groups of terms involving unknowns, or bilinear terms, and independent linear terms. These represent the integrands that need to be computed during the assembly process. Following the standard FEM procedure, the bilinear terms will correspond to assembly of elemental matrices and linear terms to elemental vectors. The details of this assembly are dependent on the code generation target. For the \texttt{Dendro} target used in this study, the assembly is done by looping over Gaussian integration points in an element and computing the integrands at those points. The following illustrates how a bilinear integrand would reflect in the generated code. Here $D$ is a coefficient, $w$ is the quadrature weight, and $J$ is a geometric factor. The basis functions and their derivatives are evaluated at the point by the functions \texttt{fe.N} and \texttt{fe.dN}. Appendix B illustrates the context for this within the generated code.\\
\begin{tabular}{r|l}
Symbolic term & $D \frac{du}{dx} v$\\
Bilinear integrand & $\frac{d\phi_{i}}{dx} * \phi_{j} * DwJ$\\
Generated code & \texttt{fe.dN(i, 0) * D * w * J * fe.N(j)}
\end{tabular}\\

In the weak form expression that is input to Finch, the separate integrals over Dirichlet and Neumann boundary regions can be easily specified by wrapping them in \texttt{dirichletBoundary(...)} and \texttt{neumannBoundary(...)} respectively. This signals to the parser that these terms represent surface integrals applied to boundary regions where those types of conditions are present. Note that in this work, the use of shifted boundary method (SBM) involves converting any Dirichlet boundary condition into equivalent Neumann conditions, so the term \texttt{dirichletBoundary} does not exactly refer to a Dirichlet condition being applied, but rather labels an integral over the boundary region where Dirichlet conditions are specified by the PDE. The relevant surface terms from Equations \eqref{eq:SBM_Dirichlet} and \eqref{eq:SBM_Neumann} (see last section) are entered into Finch as:
\begin{verbatim}
dirichletBoundary(
 -dot(grad(u), normal()) * v
 - dot(grad(v), normal())
  * (u + dot(grad(u), distanceToBoundary()) 
   - dirichletValue())
 + alpha / elementDiameter()
  * (u + dot(grad(u), distanceToBoundary()) 
   - dirichletValue()) 
  * (v + dot(grad(v), distanceToBoundary()))
)
neumannBoundary(
 dot(normal(), trueNormal()) 
  * (neumannValue() + dot(grad(u), trueNormal())) * v 
 - dot(grad(u), normal()) * v
)
\end{verbatim}

The default code generation path is for Finch's own Julia-based 
environment. It provides utilities to construct a uniformly structured
mesh and then carve out a non-conforming subdomain, such as a circle
carved from a square domain. The shifted boundary method can then be 
used by either manually specifying the integrals above, or through
an automated option. This is a good way of evaluating the method and
quickly testing variations. One can then apply this technique to
larger-scale applications using specialized frameworks external to 
the DSL.

\subsection{Generating for an external library}
One of the key features of Finch that is relevant to this study is the ability to generate code for \texttt{Dendro} as described previously. The default internal target is well suited to moderately sized problems using unstructured meshes, but for large-scale problems where more specialized tools are beneficial, external software packages are a good option. In the case of \texttt{Dendro}, we can take advantage of the sophisticated octree-based framework without investing the time to code a solution around it. As an example of this savings, the Finch script used in the demonstrations in the next section could be written in as little as 15 lines of code (see appendix A for an example). The C++ code generated by it consists of over 2500 lines of code. For more advanced users who are familiar with the target software, the generated code is also intended to be human-readable and can be easily modified as desired. All of the examples demonstrated in this paper did not require hand modification. The code was used directly as generated.

The \texttt{Dendro} target is set up to accept a wide range of input parameters that are either placed in a configuration file for the resulting program to read, or directly influence the generated code files, or both. One powerful example of this is controlling mesh refinement directly through Finch input. For example, an expression can be constructed that will be used as a criteria for refining or coarsening the mesh. As an example, the expression 
\begin{verbatim}
refineWhere="level < (sqrt(x*x+y*y)*8.2) && level < 9"
\end{verbatim}
will cause the tree refinement depth to increase with distance from the origin up to a maximum of 9. 
A simpler refinement technique that is used in this study is to set a higher depth along the boundary of the carved-out geometry. This desired depth is included as part of the geometry specification. A description of currently available input parameters is given here.

General options:
\begin{itemize}
\item baseRefineLevel: Base refinement level for full domain.
\item channelWallRefineLevel: Refinement at full domain walls.
\item refineWalls: Flags for refining at domain walls.
\item min/max: Coordinate extents for full domain.
\item matrixFree: Use a matrix-free method.
\end{itemize}

Geometry options:
\begin{itemize}
\item meshFile: File name for the mesh.
\item meshName: Label for the geometry.
\item position: Offset within the full domain for embedding the geometry.
\item outerBoundary: Does the geometry enclose the domain or exterior?
\item refineLevel: Refinement level at the edge of geometry.
\item boundaryTypes: Boundary condition type for each region.
\item bids: Boundary ID for each region.
\end{itemize}

PETSc options: Any provided PETSc KSP and PC options will be passed to PETSc to configure the linear algebra tools.

Finch also allows convenient specification of boundary regions. A user provides boolean expressions that evaluate to \texttt{true} at a given set of coordinates $(x,y,z,t)$. Since only boundary face coordinates will be input to this expression, it does not need to specify exact boundary coordinates. For example, when using a unit sphere centered at the origin, an equatorial band ranging from $z=-0.1$ to $0.1$ with boundary ID 1 can be defined by
\begin{verbatim}
addBoundaryID(1, "abs(z) <= 0.1")
\end{verbatim}
Then a Dirichlet condition for a variable $u$ that has value $exp(-\alpha z^2)$ where $\alpha=200$ could be specified by
\begin{verbatim}
alpha = 200
boundary(u,1,DIRICHLET, "exp(-$(alpha) * z^2)")
\end{verbatim}
Each of these features were employed in the demonstrations in the next section. 

\section{Demonstrations}
To demonstrate the usage and performance of these tools, we consider a time-dependent heat equation in both 2-D and 3-D. We are discretizing with a CG finite element method based on linear quadrilateral and hexahedral elements. The weak form of the equation is
\begin{equation}
\left( \frac{du}{dt}, v \right)_{\Omega} + (D\nabla u , \nabla v)_{\Omega} - (f,v)_{\Omega} - (D\nabla u \cdot \textbf{n}, v)_{\Gamma} = 0
\label{eq:heat-weakform}
\end{equation}
where $(a,b)_{\Omega}$ refers to a volume integral of $ab$ over the domain, and $(a,b)_{\Gamma}$ is a surface integral over the boundary. Here $u$ is the temperature, $D$ is the diffusion rate, and $v$ is a test function. 

Using the shifted boundary method, the boundary integral is split into Dirichlet and Neumann regions and Nitche's approximation is applied similar to \eqref{eq:SBM_Dirichlet} and \eqref{eq:SBM_Neumann}. This is the first-order formula with displacement vector $\textbf{d}$ and true normal $\textbf{n}$. The tilde refers to quantities for the surrogate boundary. The $D$ and $N$ subscripts refer to Dirichlet and Neumann boundary regions respectively.
\begin{equation}
\begin{split}
(\nabla u \cdot \textbf{n}, v)_{\Gamma} \rightarrow \int \nabla u \cdot \tilde{\textbf{n}} v d\tilde{\Gamma}_D 
\\
+ \int \nabla v \cdot \tilde{\textbf{n}} (u+\nabla u \cdot \textbf{d} -g_{D})d\tilde{\Gamma}_D 
\\
- \frac{\alpha}{h}\int (u+\nabla u \cdot \textbf{d} -g_{D})(v+\nabla v \cdot \textbf{d}) d\tilde{\Gamma}_D 
\\
+ \int (\tilde{\textbf{n}} \cdot \textbf{n})(-g_{N} - \nabla u \cdot \textbf{n}) v d\tilde{\Gamma}_{N} + \int \nabla u \cdot \tilde{\textbf{n}} v d\tilde{\Gamma}_{N}
\end{split}
\label{eq:heat-sbm}
\end{equation}

The equation is entered into Finch with the following expression, where \texttt{(...)} are the corresponding code presented in the previous section.
\begin{verbatim}
Dt(u*v) + D*dot(grad(u),grad(v)) - f*v
 + dirichletBoundary(D*...)
 + neumannBoundary(D*...)
\end{verbatim}

For these demonstrations we will set $D=1$ and $f=0$ for simplicity.

\subsection{2-D circle}
\begin{figure}[htb]
    \centering
    \includegraphics[width=\linewidth]{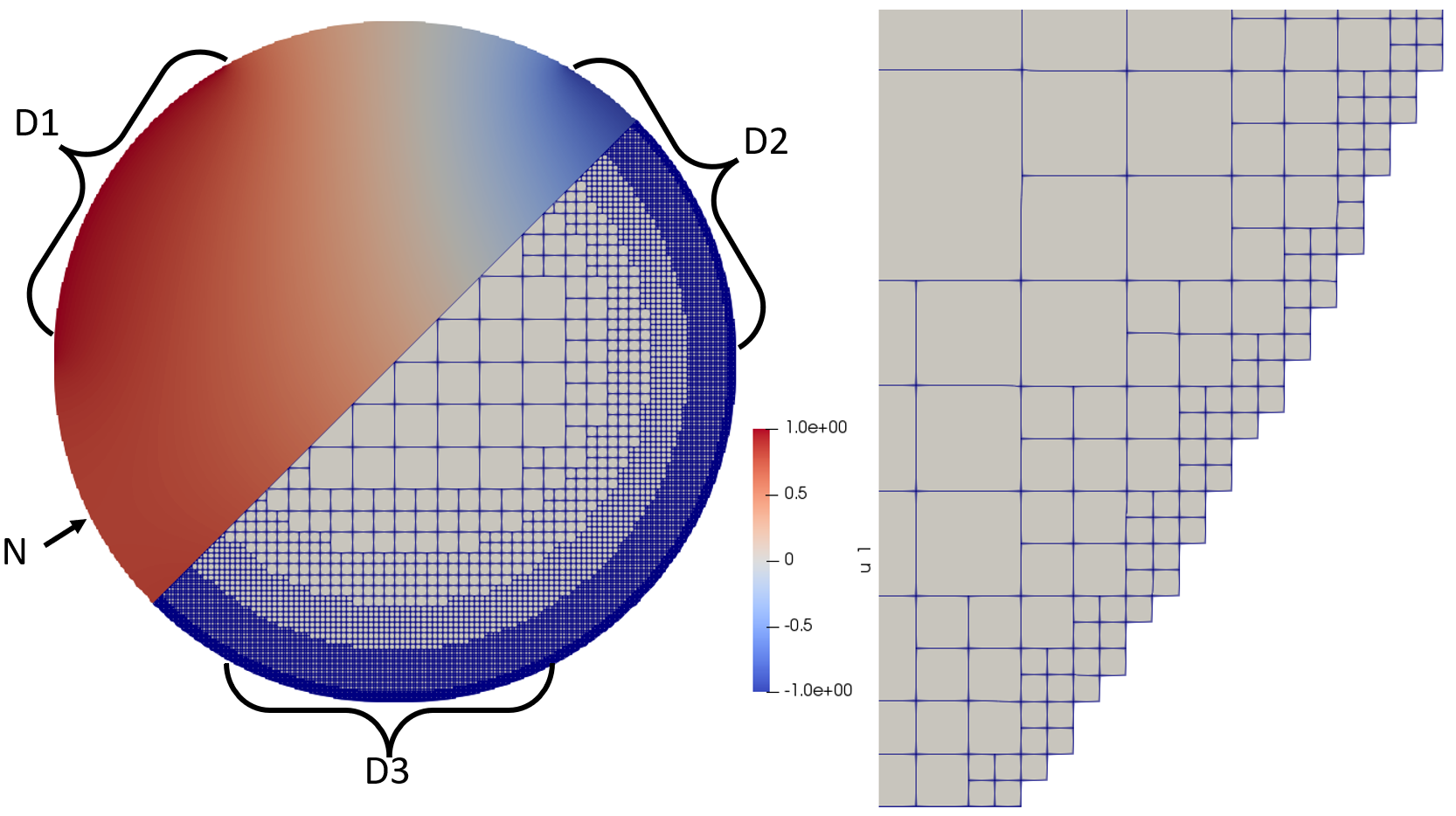}
    \caption{\textit{Left} 2-D circle mesh with various boundary regions. D1 and D3 are Dirichlet conditions with value 1, D2 is Dirichlet with value -1, and everywhere else is Neumann with value zero. Half of the figure shows the temperature value. The other half shows the refinement of the tree. \textit{Right} A close view of the edge shows the stepped pattern of the quads and the refinement along the edge.}
    \label{fig:circle}
\end{figure}

We use a 2-D circular mesh to illustrate some of the concepts. As shown in figure \ref{fig:circle}, the circle has four boundary regions. There are three sections with Dirichlet boundary conditions and the remaining boundary has Neumann conditions. This specification is made through Finch by first defining the regions.
\begin{verbatim}
# top left sixth
addBoundaryID(1,"y >= 0 && x < -0.5")
# top right sixth
addBoundaryID(2,"y >= 0 && x > 0.5")
# bottom sixth
addBoundaryID(3,"y < 0 && abs(x) < 0.5")
# everything else
addBoundaryID(4,"true")
\end{verbatim}
Note that for this target the regions can overlap. They are checked in order and the first one satisfied is selected. The next step is to define the boundary conditions for the variables in these regions. In this case we are using constants, but these could also be given as functions of position and time.
\begin{verbatim}
boundary(u,1, DIRICHLET, 1)
boundary(u,2, DIRICHLET, -1)
boundary(u,3, DIRICHLET, 1)
boundary(u,4, NEUMANN, 0)
\end{verbatim}

We also set the mesh refinement criteria through Finch input. The depth to which the tree is refined is determined by a base depth, a finer depth along the boundary of the geometry, and a custom criteria. The mesh displayed in the left, lower half of figure \ref{fig:circle} used the following criteria.
\begin{verbatim}
refineWhere="level < (sqrt(x*x+y*y) * 7.2) && level < 8"
\end{verbatim}
As long as this expression evaluates to \texttt{true} at the coordinates $(x,y)$, the mesh will be further refined. This example results in depth gradually increasing with distance from the center up to a maximum value of 8.

The geometry was provided in a Gmsh mesh file \texttt{circle.msh}, and the parameters related to this geometry are input to Finch in the form of a dictionary.
\begin{verbatim}
geometries=[Dict([
    (:meshFile, "circle.msh"),
    (:meshName, "circle"),
    (:boundaryTypes, 
        ["sbm","sbm","sbm","neumann_sbm"]),
    (:refineLevel, 7)
])]
\end{verbatim}

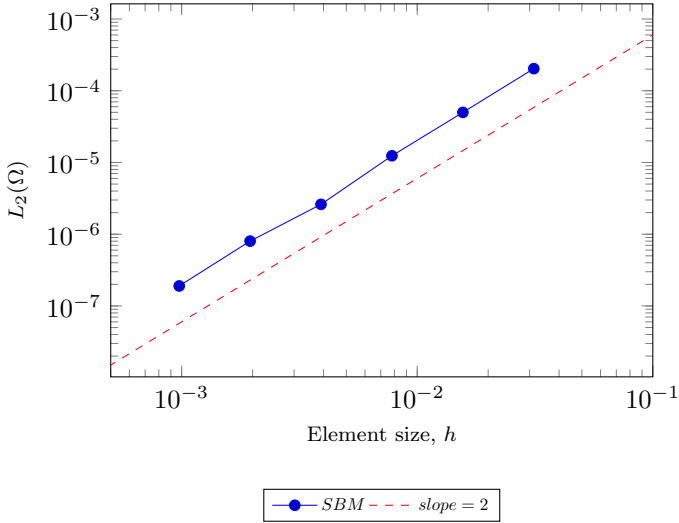
\begin{figure}[t!]
    \centering
    \begin{tikzpicture}
    \begin{loglogaxis}[
        width=0.94\linewidth, 
        height = 0.7\linewidth,
        scaled y ticks=true,
        xlabel={\footnotesize Element size, $h$},
        ylabel={\footnotesize $L_2 (\Omega)$},
        legend entries={$SBM$, $slope = 2$},
        legend style={at={(0.5,-0.3)},anchor=north, nodes={scale=0.65, transform shape}}, 
        legend columns=3,
        xmin=5e-4,
        xmax=0.1	
        ]
        \addplot table [x={h},y={L2_1p0_adaptive},col sep=comma] {figures/L2norm/L2error2DcircleOpt.csv};
        \addplot +[mark=none, red, dashed] [domain=1e-4:1]{0.06*x^2};
    \end{loglogaxis}
    \end{tikzpicture}
    \label{fig:Poisson_disk_NormL2}
\caption{Mesh convergence plot of solving the Poisson's equation on a circle.}
\label{fig:Poisson_circle}
\end{figure}

To verify the correctness of the SBM implementation, we pick up the exact solution for the circle domain to be $u(r) = 0.25(R^2-r^2)+u_0$, where $r$ is the distance from the center ($r = \sqrt{(x - x_0)^2 + (y - y_0)^2}$) and $u_0 = 0.01$. The circle has radius $R = 0.5$, and center at ($x_0 = 0.5, y_0 = 0.5$). With this analytical solution, we can find the boundary condition on the true boundary ($\Gamma$) to be $u_0 = 0.01$ and the forcing term $f = 1$. We choose the penalty parameter $\alpha$ to be 400. And, we try to perform the mesh convergence study on it with $ L_2 (\Omega) = ||u^h - u_{exact}||_{L_2(\Omega)} = \sqrt{\int_{\Omega} (u^h - u_{exact})^2 d\Omega} $ and element size $h$ ($h=1/2^l$, where $l$ ranges from 5 to 10). We observe the ideal slope of the convergence rate with the order of accuracy close to 2, shown in Figure \ref{fig:Poisson_circle}.

\subsection{3-D irregular meshes}
\begin{figure*}[bpt]
    \centering
    \includegraphics[width=0.8\linewidth]{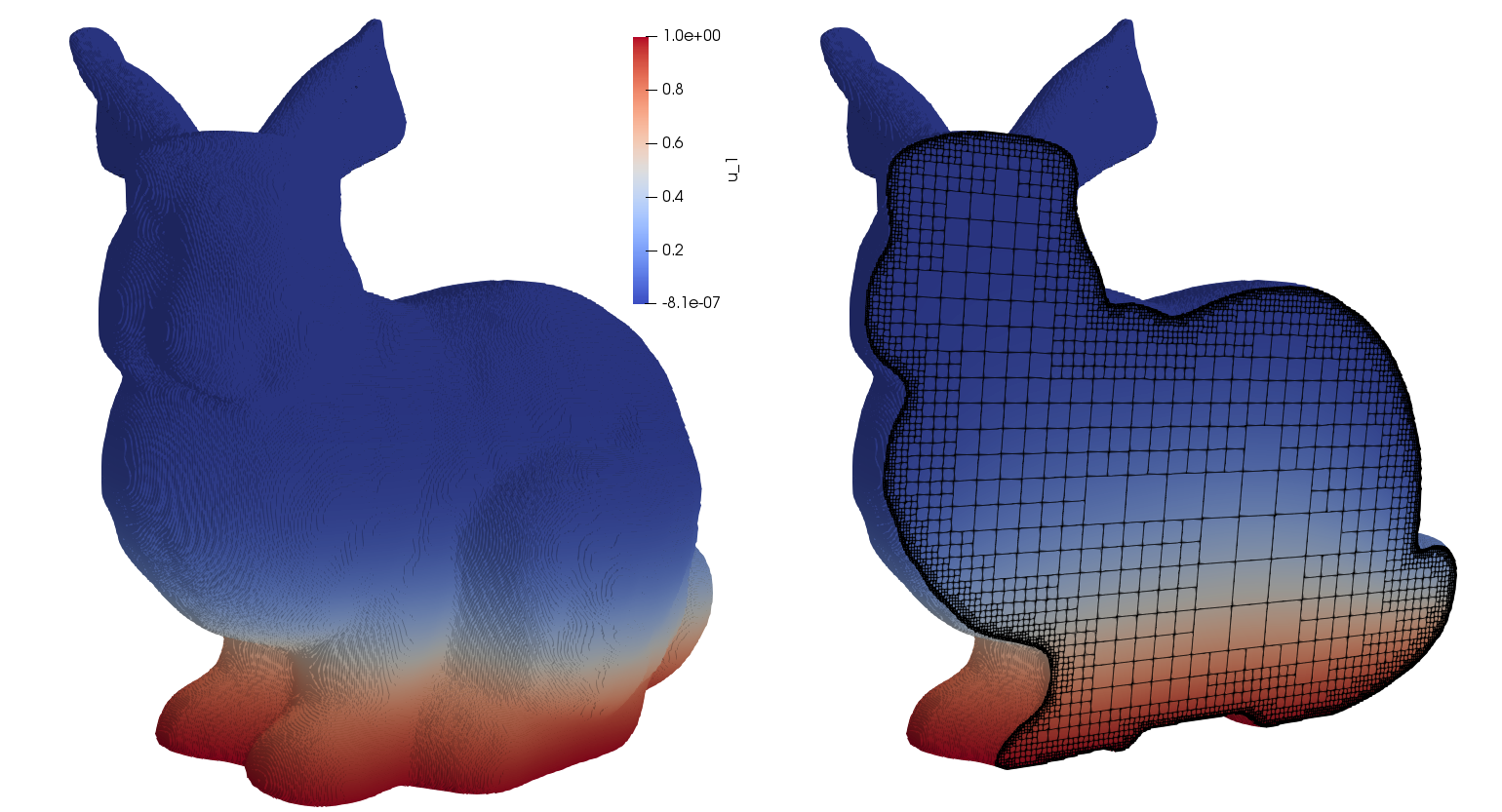}
    \caption{\textit{Left} Solution using the bunny mesh. \textit{Right} A cut through the center showing the octree structure that is refined along the boundary. The interior was kept coarse to emphasize the refinement.}
    \label{fig:bunny}
\end{figure*}

\begin{figure*}[pbt]
    \centering
    \includegraphics[width=0.8\linewidth]{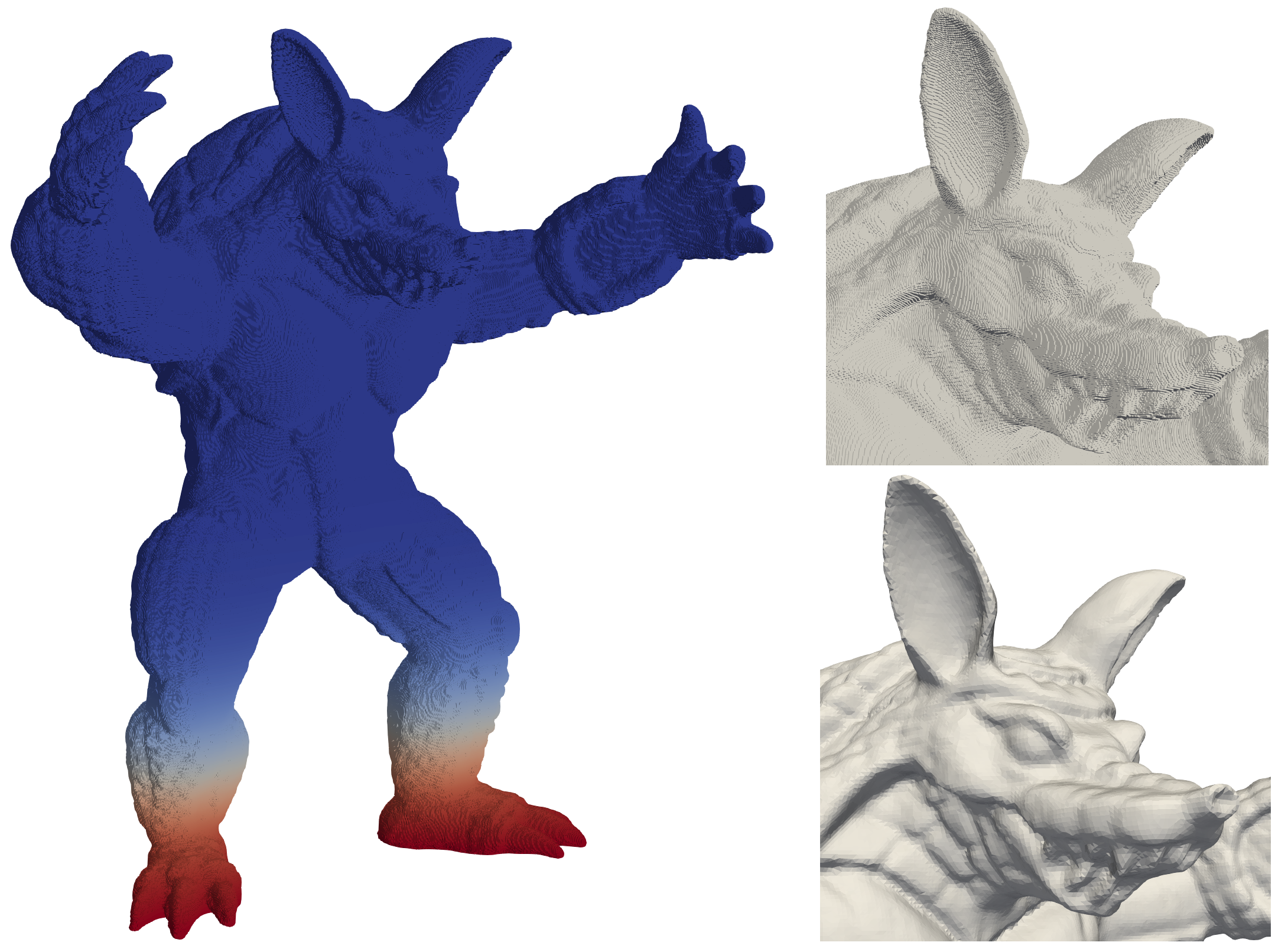}
    \caption{\textit{Left} Solution using the armadillo mesh. \textit{Upper Right} A close view of the face to show the non-conforming set of octants. \textit{Lower Right} A similar view of the original mesh surface.}
    \label{fig:arma}
\end{figure*}

\begin{figure*}[pbt]
    \centering
    \includegraphics[width=0.75\linewidth]{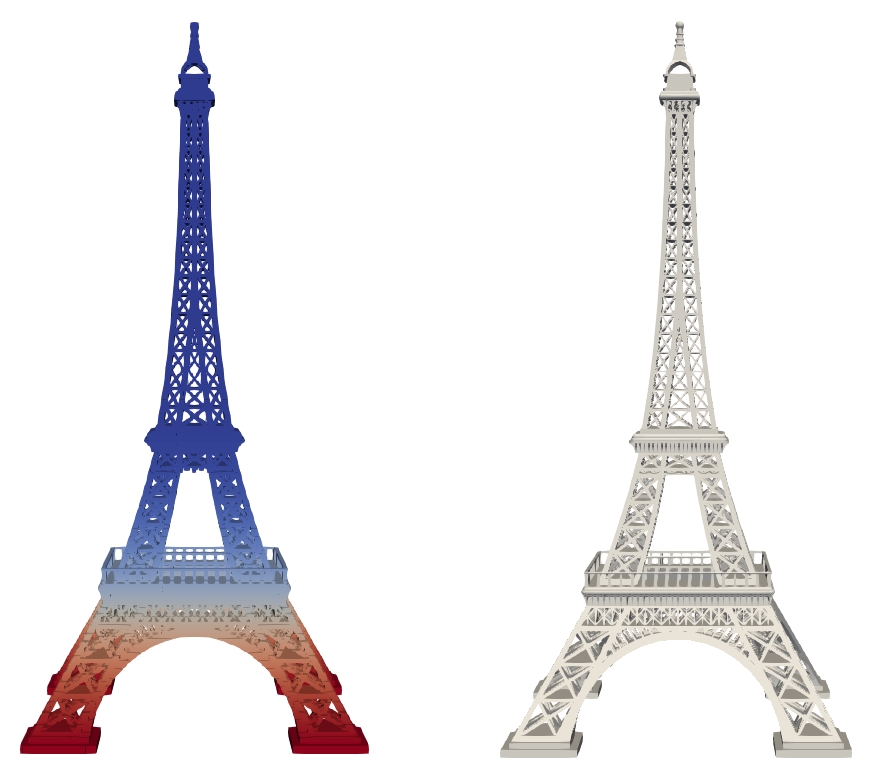}
    \caption{\textit{Left}An Eiffel tower mesh illustrating a fine structure that is well represented by this set of 2.8 million octants. \textit{Right} The original mesh.}
    \label{fig:eiffel}
\end{figure*}

Several 3-D meshes were given a similar treatment based on the heat equation. Refer to figures \ref{fig:bunny}, \ref{fig:arma}, and \ref{fig:eiffel}. These examples were all given zero initial temperature and a warmer Dirichlet condition near the base. Figure \ref{fig:bunny} includes a cut-away to illustrate the refinement of the mesh along the boundary in order to approximate the smooth background geometry.

This approximation is apparent in figure \ref{fig:arma} when looking closely at intricately detailed parts of the geometry. Despite the roughness of the octants, the shifted boundary method does give a boundary condition that approximates the condition on the true boundary rather than the imprecise surrogate boundary of the octants.

The ability to capture very intricate detail and complexity of the true geometry is demonstrated in figure \ref{fig:eiffel}. As long as the refinement depth is chosen large enough, an arbitrarily detailed input mesh can be well approximated.

To test whether the SBM works well in solving these intricate geometries, we conduct a mesh convergence study on a 3D Stanford bunny. Specifically, we employ a manufactured solution of the form:
\begin{equation}  \label{eq:Solution_bunny}
    u(x,y,z) = cos(\pi x) y sin(\pi z).
\end{equation}

The $f$ corresponding to the manufactured solution is 
\begin{equation}  \label{eq:f_bunny}
    2 \pi^2 cos(\pi x) y sin(\pi z).
\end{equation}

\begin{figure}[t!]
    \centering
	\begin{tikzpicture}
	\begin{loglogaxis}[ width=0.94\linewidth, height = 0.7\linewidth,, scaled y ticks=true,xlabel={Element size, h},ylabel={\footnotesize $L_2 (\Omega)$},legend entries={SBM, $slope = 2$},
	legend style={at={(0.5,-0.25)},anchor=north, nodes={scale=0.65, transform shape}}, 
	legend columns=2,
	xmin=3e-3,xmax=1e-1	]
	\addplot table [x={h},y={L2_1p0_adaptive},col sep=comma] {figures/L2norm/L2error3DbunnyOpt.csv};
	\addplot +[mark=none, red, dashed] [domain=0.001:1]{0.1*x^2};
	\end{loglogaxis}
	\end{tikzpicture}
 \caption{Mesh convergence plot of solving the Poisson's equation on 3D Stanford bunny.}
\label{fig:Poisson_bunny}
\end{figure}
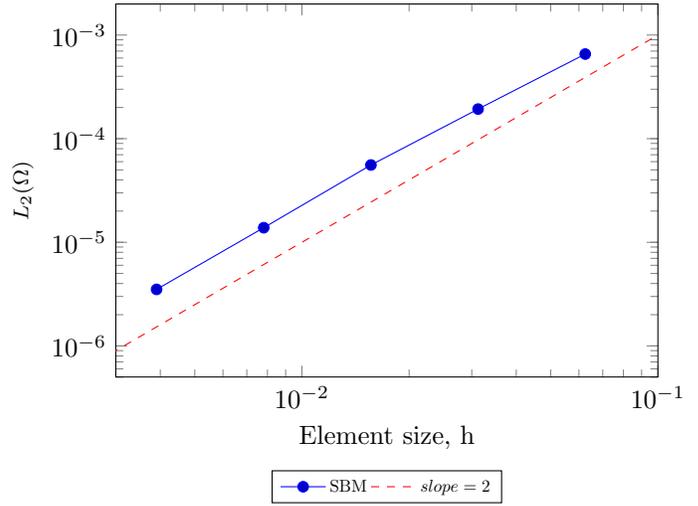

The boundary condition we enforce on the true boundary ($\Gamma$) through SBM is $u_D = cos(\pi x) y sin(\pi z)$ and the element size we set is $h=1/2^l$, where $l$ ranges from 4 to 8. Again, we obtain a nice convergence rate close to a slope of 2, shown in Figure \ref{fig:Poisson_bunny}, which demonstrates the robustness of SBM to deal with complicated 3D objects.

\subsection{Performance}
To measure parallel performance of the code, the above setup was used with the Stanford bunny mesh at a variety of refinements and processor counts. The code was run on the Frontera supercomputer at the Texas Advanced Computing Center. The hardware consists of  dual socket Intel Xeon Platinum 8280 nodes with 56 cores and 192GB of RAM per node. The PETSc options used were:
\begin{itemize}
\item ksp\_type = bcgs
\item ksp\_max\_it = 1000
\item ksp\_atol = 1e-8
\item ksp\_rtol = 1e-8
\item pc\_type = bjacobi
\end{itemize}

The refinements used were 1) base depth = 4, boundary depth = 7, which resulted in 107,323 nodal degrees of freedom, 2) base = 5, boundary = 9, with 1,777,754 degrees of freedom, and 3) base = 6, boundary = 11, with 28,729,810 degrees of freedom.

\begin{figure}[!hbt]
    \centering
    \includegraphics[width=\linewidth]{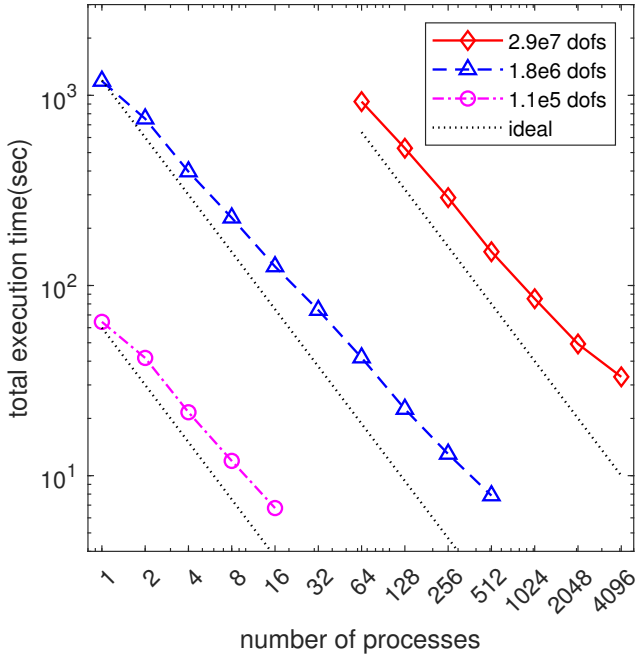}
    \caption{Strong scaling for three different refinements with the heat equation. dofs refers to the number of nodal degrees of freedom.}
    \label{fig:scaling}
\end{figure}

\begin{figure}[!hbt]
    \centering
    \includegraphics[width=\linewidth]{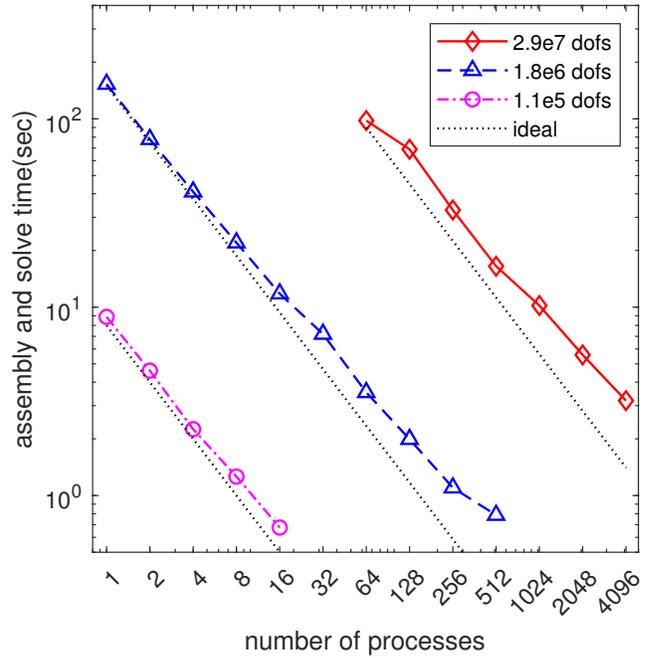}
    \caption{Strong scaling similar to figure \ref{fig:scaling}, but excluding the octree generation portion.}
    \label{fig:solvescaling}
\end{figure}

Figure \ref{fig:scaling} shows the strong scaling for the three refinements over a wide range of process counts. Good scaling was achieved for all cases. In particular, the finest refinement displayed very consistent scaling up to several thousand processes. Figure \ref{fig:solvescaling} is similar, but excludes the octree generation portion of the work. This shows that the assembly and solve portions also scale well. These portions are particularly related to the generated code.

\begin{figure}
    \centering
\begin{tikzpicture}

  \begin{axis}[
    ybar stacked, ymin=0,  
    bar width=3mm,
    symbolic x coords={64,128,256,512,1024,2048,4096},
    xtick=data,
    xlabel=Number of processes,
    ylabel=percentage of execution time,
    legend columns=2,
    legend style={  at={(0.5,1)},
        /tikz/column 2/.style={
            column sep=5pt},
        anchor=south}
  ]
  
  \addplot [fill=blue] coordinates {
({64},88.98638767)
({128},86.38070765)
({256},88.19294456)
({512},88.50174216	)
({1024},87.48466258)
({2048},88.24768324)
({4096},89.73946607)};
\addlegendentry{octree generation}
  \addplot [fill=orange] coordinates {
({64},2.902463719)
({128},2.80687883)
({256},2.796976242)
({512},2.696864111)
({1024},2.429447853)
({2048},2.274641955)
({4096},2.071405597)};
\addlegendentry{vector assembly}
  \addplot [fill=green] coordinates {
({64},3.521205985)
({128},3.340581143)
({256},3.318934485)
({512},3.219512195)
({1024},3.042944785)
({2048},2.843302443)
({4096},2.894821486)};
\addlegendentry{matrix assembly}
  \addplot [fill=yellow] coordinates {
({64},4.589942626)
({128},7.471832378)
({256},5.691144708)
({512},5.581881533)
({1024},7.042944785)
({2048},6.634372367)
({4096},5.294306851)};
\addlegendentry{linear solve}
  
  \end{axis}
  
\end{tikzpicture}
    \caption{Percentage of total computation time spent on major steps. This is 
    for the finest refinement on the bunny mesh from the scaling results above.}
    \label{fig:percents}
\end{figure}
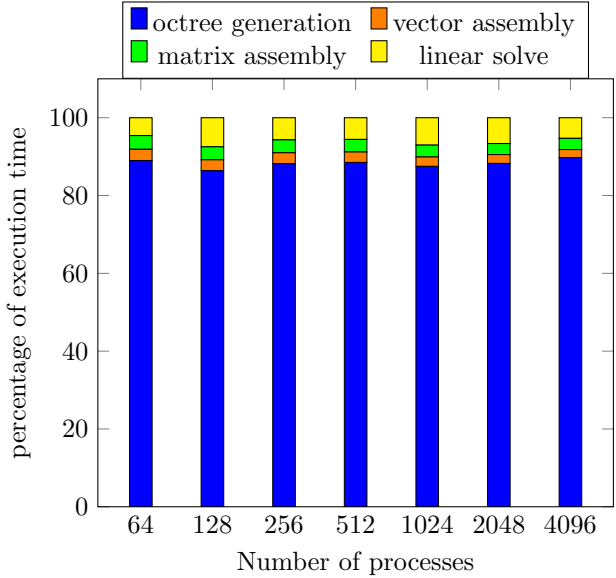

The computation was then profiled using PETSc's profiling tool. Figure \ref{fig:percents} shows the percentage of execution time spent in octree generation, vector assembly, matrix assembly, and linear system solve. These results correspond to the scaling experiment described above for the finest refinement. 
It is immediately clear that a large percentage of time is spent on mesh creation. The portions for vector and matrix assembly are similar, and the linear solve is slightly longer than the assembly. Note that the proportions stay roughly equal for all numbers of processes. This suggests that each part scales similarly.

It is also worth noting that this configuration has a significantly higher refinement depth along the boundary compared to the base depth on the interior of the bunny. The assembly is more expensive along the boundary due to the integrals introduced by the shifted boundary method as well as the displacement vector search.This means that a large portion of the degrees of freedom are on the boundary and require the costly displacement vector search. 


\section{Shifted boundary method: Weak formulations for various PDEs}

Here we introduce SBM applied to three different PDEs. We talk about the Poisson equation first.
For example, for enforcing the Dirichlet boundary condition ($u = u_D$) using SBM, we can write the weak form of the Poisson equation as:
\begin{equation}
\begin{split}
    &(\nabla w^h, \nabla u^h)_{\Tilde{\Omega}} - (w^h,\nabla u^h \cdot \Tilde{\mathbf{n}})_{\Tilde{\Gamma}} \\
    -& (\nabla w^h \cdot \Tilde{\mathbf{n}}, u^h + \nabla u^h \cdot \mathbf{d} -u_D)_{\Tilde{\Gamma}}  \\
    +& \frac{\alpha}{h}(w^h + \nabla w^h \cdot \mathbf{d}, u^h + \nabla u^h \cdot \mathbf{d} - u_D)_{\Tilde{\Gamma}} = (w^h,f)_{\Tilde{\Omega}} ,
\end{split}
\label{eq:SBM_Dirichlet}
\end{equation}
where $\Tilde{\Omega}$ is the surrogate domain, $\Tilde{\Gamma}$ is the surrogate boundary, $\Tilde{n}$ is the normal vector of the integration point on the surrogate boundary, $\alpha$ is the penalty parameter for the Dirichlet boundary condition, $h$ is the element size, and $\textbf{d}$ is the displacement vector from the surrogate boundary $\Tilde{\Gamma}$ to the true boundary $\Gamma$. 

In addition to the Dirichlet boundary condition, in order to apply the Neumann boundary condition ($-\frac{\partial u}{\partial n} = t_N$) to the Poisson problem, the formulation is written as:
\begin{equation}
\begin{split}
    (\nabla w^h, \nabla u^h)_{\Tilde{\Omega}} - (w^h,\nabla u^h \cdot \Tilde{\mathbf{n}})_{\Tilde{\Gamma}} + (w^h,(\Tilde{\mathbf{n}} \cdot \mathbf{n})(t_N + \nabla u \cdot n))_{\Tilde{\Gamma}}  \\ 
= (w^h,f)_{\Tilde{\Omega}} ,
\end{split}
\label{eq:SBM_Neumann}
\end{equation}
where \textbf{n} is the true normal vector of the projection point on the true boundary $\Gamma$ associated with the integration point on the surrogate boundary $\Tilde{\Gamma}$. 

SBM can also be extended to linear elasticity. The weak formulation for static linear elasticity with $\boldsymbol{u} = \boldsymbol{u_D}$ as the Dirichlet boundary condition, where  $\boldsymbol{u}$ is the displacement field, can be written as:
\begin{equation}
\begin{split}
&(\nabla^s \mathbf{w^h}, \boldsymbol{C} \boldsymbol{\varepsilon}(\boldsymbol{u^h}) )_{\Tilde{\Omega}} - (\mathbf{w^h},(\boldsymbol{C} \boldsymbol{\varepsilon}(\boldsymbol{u^h})) \cdot \Tilde{\mathbf{n}})_{\Tilde{\Gamma}}  \\
+ &((\mathbf{C}\nabla^s \mathbf{w^h}) \cdot \Tilde{\mathbf{n}}, \boldsymbol{u^h} + \nabla \boldsymbol{u^h} \cdot \mathbf{d} -\boldsymbol{u_D})_{\Tilde{\Gamma}} \\
+ &\frac{\gamma}{h}(\mathbf{w^h} + \nabla \mathbf{w^h} \cdot \mathbf{d}, \boldsymbol{u^h} + \nabla \boldsymbol{u^h} \cdot \mathbf{d} -\boldsymbol{u_D})_{\Tilde{\Gamma}} = (\mathbf{w^h},\mathbf{b})_{\Tilde{\Omega}} .
\end{split}
\label{eq:SBMLE}
\end{equation}
 where $\mathbf{C}$ is the elastic stiffness tensor, $\nabla^s$ is the symmetric gradient operator, $\boldsymbol{\varepsilon}$ is the strain tensor, $\mathbf{b}$ is the body force, and $\gamma$ is the penalty parameter for the Dirichlet boundary condition of the linear elasticity. 
 
 Finally, the SBM surface integration terms to enforcing velocity field $\boldsymbol{u} = \boldsymbol{u_D}$ as the Dirichlet boundary condition for the non-dimensionalized Navier-Stokes equation can be expressed as:
 \begin{equation}
\begin{split}
  -&(\mathbf{w^h},(\frac{2}{Re} \boldsymbol{\varepsilon}(\boldsymbol{u^h}) - p^h) \cdot \Tilde{\mathbf{n}})_{\Tilde{\Gamma}} \\
  -&((\frac{2}{Re} \nabla^s \mathbf{w^h} + q^h) \cdot \Tilde{\mathbf{n}},\boldsymbol{u^h} + \nabla \boldsymbol{u^h} \cdot \mathbf{d} -\boldsymbol{u_D})_{\Tilde{\Gamma}}  \\
  + &\frac{1}{Re} \frac{\beta}{h}(\mathbf{w^h} + \nabla \mathbf{w^h} \cdot \mathbf{d}, \boldsymbol{u^h} + \nabla \boldsymbol{u^h} \cdot \mathbf{d} -\boldsymbol{u_D})_{\Tilde{\Gamma}}
\end{split}
\label{eq:SBMNS}
\end{equation}
where $p^h$ is the pressure field we want to solve for, Re is the Reynolds number, $\boldsymbol{\varepsilon}$ is the strain rate tensor, and $\beta$ is the penalty parameter for the Dirichlet boundary condition of the Navier-Stokes equation. 

\section{Conclusion}
We have presented a set of software tools for computing finite element solutions to PDEs based on a scalable, parallel $k$-d tree library driven through a high-level domain specific language. The shifted boundary method based on Nitche's method was employed to allow this structured mesh technique to handle boundaries of non-conforming, irregular geometries.

This method requires a complicated set of boundary integrals that can be very easily specified in the DSL. Other aspects, such as setting up complex boundary regions and defining mesh refinement criteria, are also performed through the intuitive interface. 

We demonstrated the scalable performance of the generated code on up to 4096 processes for about 29 million degrees of freedom. An analysis of computational time revealed that the linear assembly and solution steps performed well and the octree generation required a significant majority of the execution time. We are currently expanding the capabilities of this code generation target to handle a wider range of applications including nonlinear equations.

\section*{Acknowledgement}
This work was supported by grants that are withheld from this review copy to maintain anonymity.

\bibliographystyle{ACM-Reference-Format}
\bibliography{ref-base}

\appendix
\section{Example Input Code}
Below is an example of the Julia input code for Finch. This is a simplified working example that is capable of generating the results presented. The forcing term is set to zero in this example.

\begin{table}[htb]
\textbf{Example input code}
\begin{verbatim}
#=
# This is a 3-D time-dependent heat equation 
# with Dirichlet boundary conditions.
# Linear, CG elements are used.
# This is a simplified example that makes use of
# the default PETSc options.
=#

using Finch # Load the package
initFinch("heat3d"); # Initialize

# Set target and corresponding options.
generateFor("dendrite", baseRefineLevel=5, 
    min=[0.0,0.0,0.0], max=[1.0,1.0,1.0], 
    geometries=[Dict([(:meshFile, "bunny.stl"), 
    (:boundaryTypes, ["sbm"]), (:refineLevel, 8)])])
domain(3) # 3-dimensional domain
timeStepper(BDF2) # BDF-2 time stepper
setSteps(0.01, 100) # Set step size and count 

# Create variables and other entities.
u = variable("u")
testSymbol("v")
coefficient("alpha", 200) # penalty parameter

# Define boundary regions. 
# Set boundary and initial conditions.
addBoundaryID(1, "true") # region 1 = full boundary
boundary(u, 1, DIRICHLET, "exp(-z*z / 0.04)")
initial(u, 0.0) # initial condition

# Write the PDE with SBM boundary integrals
weakForm(u, "Dt(u*v) + dot(grad(u),grad(v)) + 
    dirichletBoundary(
        -dot(grad(u), normal()) * v
        - dot(grad(v), normal())
        * (u + dot(grad(u), distanceToBoundary()) 
        - dirichletValue())
        + alpha / elementDiameter() 
        * (u + dot(grad(u), distanceToBoundary())
        - dirichletValue())
        * (v + dot(grad(v), distanceToBoundary())))")

solve(u);
\end{verbatim}
\end{table}

\section{Excerpt of generated code}
This sample of the generated code performs the quadrature
for the elemental matrix and vector assembly.

\begin{table}
\textbf{Example generated code excerpt} 
\begin{verbatim}
/* Volume integrals/////////////////////////////
This is called once for each quadrature point.
Will be inside this loop:
while (fe.next_itg_pt()) {
	Integrands_Ae(fe, Ae);
}
*/
void Integrands_Ae(const FEMLIB::FEMElm &fe, 
            FEMLIB::ZeroMatrix<double> &Ae) {
  using namespace FEMLIB;
  // # of dimensions: 1, 2, or 3
  const int n_dimensions = 3;
  // # of basis functions
  const int n_basis_functions = fe.nbf();

  // (determinant of J) cross W
  const double wdetj = fe.detJxW();
  // coordinates of this point
  const ZEROPTV p = fe.position();

  for (int row = 0; row < n_basis_functions; row++) {
    for (int col = 0; col < n_basis_functions; col++) {
      ////////////////////////////////////////////
      // This is generated from the input expressions
      double N = 0.0;
      N += (fe.N(row)*(wdetj * 1.5)*fe.N(col));
      N += (fe.dN(row, 0)*(wdetj * dt)*fe.dN(col, 0));
      N += (fe.dN(row, 1)*(wdetj * dt)*fe.dN(col, 1));
      N += (fe.dN(row, 2)*(wdetj * dt)*fe.dN(col, 2));
      Ae(row, col) += N;
      ////////////////////////////////////////////
    }
  }
}
void Integrands_be(const FEMLIB::FEMElm &fe, 
            FEMLIB::ZEROARRAY<double> &be) {
  using namespace FEMLIB;
  // # of basis functions
  const int n_basis_functions = fe.nbf();
  // (determinant of J) cross W
  const double wdetj = fe.detJxW();
  // coordinates of this point
  const ZEROPTV p = fe.position();

  // Evaluate coefficients for volume vector. //
  double value__u_1 = p_data_->valueFEM(fe, 0);
  double value_PREV2_u_1 = 
            p_data_->valueFEM(fe, 0 + NUM_VARS);

  for (int row = 0; row < n_basis_functions; row++) {
    ///////////////////////////////////////////////
    // This is generated from the input expressions
    double N = 0.0;
    N += (fe.N(row)*(wdetj*(2 * value__u_1)));
    N += (fe.N(row)*(wdetj*(-0.5 * value_PREV2_u_1)));
    be(row) += N;
    ///////////////////////////////////////////////
  }
}
\end{verbatim}
\end{table}

\end{document}